\documentclass{article}
\usepackage[totalwidth=480pt, totalheight=680pt]{geometry}

\usepackage{graphicx}
\usepackage{natbib}
\usepackage{amsmath}

\graphicspath{{figs/}}

\title{Localised plumes in three-dimensional compressible magnetoconvection}
\author{S.~M. Houghton$^1$
\thanks{Email: smh@maths.leeds.ac.uk} 
and P.~J. Bushby$^2$
\thanks{Email: paul.bushby@ncl.ac.uk} \\
$^1$ School of Mathematics, University of Leeds, LS2 9JT, UK.\\
$^2$ School of Mathematics \& Statistics, Newcastle University, Newcastle upon Tyne, NE1 7RU, UK.}

\begin{document}

\date{Draft prepared \today}

\maketitle

\label{firstpage}

\begin{abstract}
Within the umbrae of sunspots, convection is generally inhibited by
the presence of strong vertical magnetic fields. However, convection
is not completely suppressed in these regions: bright features, known
as umbral dots, are probably associated with weak, isolated convective
plumes. Motivated by observations of umbral dots, we carry out
numerical simulations of three-dimensional, compressible
magnetoconvection. By following solution branches into the subcritical
parameter regime (a region of parameter space in which the static
solution is linearly stable to convective perturbations), we find that
it is possible to generate a solution which is characterised by a
single, isolated convective plume. This solution is analogous to the
steady magnetohydrodynamic convectons that have previously been found in
two-dimensional calculations. These results can be related, in a
qualitative sense, to observations of umbral dots.   
\end{abstract}


\section{Introduction}
Using modern instruments, such as the Solar Optical Telescope on
board Hinode and the 1-metre Swedish Solar Telescope on La
Palma, it is possible to make detailed observations of magnetic fields
and convection at the surface of the Sun. Sunspots are the most
prominent magnetic features on the solar surface. A typical sunspot
consists of a central umbral region, surrounded by a complex filamentary
penumbra. Umbral regions appear dark because their surface
temperatures are (typically) only $70-85\%$ of the mean surface
temperature of the non-magnetic photosphere  \citep[see, for
example,][]{TW}. This reduction in temperature is due to the fact that
the convective transport of heat is impeded within sunspot umbrae by
the presence of strong, near vertical magnetic fields (which can often
exceed $3000$G).   

Detailed observations of sunspot umbrae have shown that they are not
uniformly dark. In almost all sunspots, bright point-like structures
can be observed -- these are known as umbral dots \citep{dan64}. These
bright features are warmer than their immediate surroundings, but are
(generally) cooler than the surrounding photosphere \citep[see, for
example,][]{sob05,Kitai}. It is difficult to determine the characteristic
size of an umbral dot, although these features are always small
compared to the umbral diameter. In a recent study, \citet{Kitai}
found that the umbral dots in one particular sunspot had typical
diameters of approximately $220-350$km, although a significant number
appeared to be much smaller than this (possibly below the resolution
limit for the Solar Optical Telescope on Hinode). Umbral dots are also
short-lived features. \citet{Kitai} found that most of the umbral dots
in their survey had lifetimes of between $5$ and $20$ minutes. In an
earlier study, \citet{sob97} found a much broader range of lifetimes
for umbral dots (with a small percentage lasting longer than $2$
hours), although, like \citet{Kitai}, they found a mean lifetime of
approximately $15$ minutes. Most umbral dots exhibit no systematic proper
motions. However, those that appear to form at the umbral/penumbral
boundary (which are often associated with penumbral grains) tend to
migrate radially inwards towards the centre of the umbra
\citep[]{sob95,Kitai}. There is some observational evidence for weak
upflows within umbral dots \citep[][]{Socas,bjj07} as well as
downflows around their edges \citep{bjj07,ort10}. Clearly, the
observations 
indicate that umbral dots correspond to convective plumes within
sunspot umbrae. Further theoretical support for this conclusion comes
from the work of \citet{dein65}, who determined that convective
motions must be present within the umbra, as radiative processes alone
could not transport sufficient energy to the surface.  

Theoretical studies of umbral convection tend to be based upon local
models of magnetoconvection in a Cartesian domain. It is well-known
that a strong vertical magnetic field tends to inhibit convective
motions in an electrically-conducting fluid \citep{chand}. When the
dynamics are dominated by magnetic fields, convection takes the form
of weak, narrow plumes. In an idealised model of magnetoconvection,
\citet{WPB} found a steady, almost hexagonal pattern of convection in
the magnetically-dominated regime. More recently, \citet{sv06}
\citep[see also][]{barti10} have carried out a more realistic set of
calculations, including the effects of partial ionisation and
radiative transfer. These simulations produced a time-dependent pattern
of individual convective plumes, the properties of which compare very
favourably to observations of umbral dots. In the calculations of
\citet{WPB} and \citet{sv06}, convective features tend to be
distributed across the whole computational domain. It is worth noting
that observations indicate that the distribution of umbral dots tends
to be rather non-uniform \citep{sob05}. This could simply be explained
by variations in intensity \citep[as seen in the calculations
of][]{sv06}, but some degree of localisation in the distribution of
convective plumes could also help to explain the observed distribution
of umbral dots. 

In a two-dimensional model of incompressible magnetoconvection,
\citet{b99} found strongly localised, steady convective states \citep[see also][]{d07}. These
were named convectons. Restricting attention to a simplified model, in
which the governing equations were projected onto a minimal set of
Fourier modes in the vertical direction, \citet{bw02} were also able
to find oscillatory localised states in three spatial dimensions. In
both these cases, these localised states were found in the subcritical
parameter regime (in which the static, purely conducting state is linearly
stable to convective perturbations). These localised states are an
extreme example of a phenomenon that is known as flux separation
\citep{twbp98}. Convective plumes tend to expel magnetic flux
\citep{w66}, causing it to accumulate in the surrounding fluid. In
this subcritical parameter regime, the magnetic field that surrounds
the plume becomes sufficiently strong that convection in the
  surrounding fluid is completely inhibited, giving rise to a truly localised convective
state. If conditions are appropriate for subcritical convection within
sunspots (something that is certainly plausible), these results
suggest that it may be possible for truly localised umbral dots to
form within sunspot umbrae.   

In this paper, we demonstrate the existence of steady localised convective
plumes in three-dimensional compressible magnetoconvection. Unlike
\citet{bw02}, we make no simplifying assumptions regarding the
vertical structure of the convective flows. The setup of the model is
described in detail in the next section of the paper. Numerical
results from this model are presented in Section 3. In the final
section, we relate our findings (in qualitative terms) to observations
of umbral dots. 

\section{Problem description and setup}\label{sec:setup}

We consider the evolution of a layer of compressible,
electrically-conducting fluid, heated from below, in the presence of
an imposed magnetic field.  Various properties of the fluid, including
the thermal conductivity, $K$, the shear viscosity, $\mu$, the magnetic
diffusivity, $\eta$, the magnetic permeability, $\mu_0$, and the
specific heat capacities at constant pressure and density ($c_P$ and
$c_V$ respectively) are assumed to be constant.  At a position
$\mathbf{x}$ and time $t$, we define $\rho(\mathbf{x},t)$,
$T(\mathbf{x},t)$ and $\mathbf{u}(\mathbf{x},t)$ to be the fluid
density, temperature and velocity field (respectively), whilst
$\mathbf{B}(\mathbf{x},t)$ represents the magnetic field.   

This fluid occupies a three-dimensional Cartesian domain
with $0 \le z \le d$ and $0\le x,\,y \le 8d$.  The axes of this
coordinate system are orientated so that the $z$-axis points
vertically downwards, parallel to the constant gravitational
acceleration, $\mathbf{g}=g\mathbf{\hat{z}}$.  For this model problem, periodic
boundary conditions are imposed in the $x$ and $y$ directions, whilst
the upper and lower boundaries (at $z=0$ and $z=d$) are assumed to be
impermeable and stress free.  Furthermore, fixed temperature boundary
conditions are applied at the upper and lower boundaries with $T=T_0$
at $z=0$ and $T=T_0+\Delta T$ at $z=d$ ($\Delta T>0$).  It is also
assumed that the horizontal components of any magnetic fields that are
present vanish at $z=0$ and $z=d$.  When the layer is static, 
the imposed magnetic field is uniform and vertical,
i.e. $\mathbf{B}=B_0\mathbf{\hat{z}}$. 

Before writing down the governing equations for this system, we
can express these in non-dimensional form.  More details of this procedure
can be found in \citet{matt} and \citet{bh05}.  Very briefly, all lengths are
scaled by the layer depth, $d$, whilst an acoustic time-scale,
$d/\left(R_*T_0\right)^{1/2}$ (where $R_*$ is the gas constant) is
used to rescale time.  After rescaling $\rho$, $T$, $\mathbf{u}$ and
$\mathbf{B}$ in an appropriate way (see \citet{bh05} for more details),
the governing equations for this system can be written in the following form:

\begin{align}
\frac{\partial \rho}{\partial t} &= {} - \nabla \cdot \left(\rho
\mathbf{u}\right),\\
\begin{split}
\frac{\partial}{\partial t}\left(\rho \mathbf{u}\right) &= {} - \nabla
\left(\rho T + F|\mathbf{B}|^2/2\right)
+\theta(m+1)\rho\mathbf{\hat{z}} \\ 
& \quad {} + \nabla \cdot \left( F \mathbf{BB} - \rho \mathbf{u u} +
\kappa \sigma \mathbf{\tau}\right),
\end{split} \\
\frac{\partial \mathbf{B}}{\partial t} &= \nabla \times \left( \mathbf{u}
\times \mathbf{B} -  \kappa \zeta_0 \nabla \times \mathbf{B} \right),\\
\begin{split}
\frac{\partial T}{\partial t} &=  {} - \mathbf{u}\cdot\nabla T - \left(\gamma
-1\right)T\nabla \cdot \mathbf{u} + \frac{\kappa\gamma}{\rho}\nabla^2
T \\
& \quad {} + \frac{\kappa(\gamma-1)}{\rho}\left(\sigma \tau^2/2 +
F\zeta_0|\nabla \times \mathbf{B}|^2\right),
\end{split}
\end{align}
where the components of the stress-tensor, $\mathbf{\tau}$, are given by 
\begin{equation}
\tau_{ij}= \frac{\partial u_i}{\partial x_j}+\frac{\partial
  u_j}{\partial x_i} - \frac{2}{3}\frac{\partial u_k}{\partial
  x_k}\delta_{ij},
\end{equation}
\noindent whilst $\mathbf{B}$ satisfies the standard constraint that
$\nabla\cdot\mathbf{B}=0$. The pressure $P$ satisfies 

\begin{equation}
P=\rho T.
\end{equation}

\noindent Equation (1) describes the conservation of
mass, whilst Equation (2) is simply the momentum equation, written in
conservative form.  Note that the $\theta(m+1)\rho\mathbf{\hat{z}}$
term in Equation (2) represents the effects of gravity whilst the two
terms that are quadratic in $\mathbf{B}$ correspond to the Lorentz
force.  The evolution of the magnetic field is governed by the
standard magnetic induction equation (3).  The final two terms in the
thermal equation (4) represent the effects of viscous and ohmic
heating.  The non-dimensional parameters that appear in the governing
equations are defined in Table 1.  For later convenience, we also
introduce the Chandrasekhar number,
\begin{equation}
Q=F/\kappa^2\zeta_0\sigma,
\end{equation}
and the Rayleigh number,
\begin{equation}
Ra=(m+1-m\gamma)(1+\theta/2)^{2m-1}\frac{(m+1)\theta^2}{\kappa^2\gamma\sigma}.
\end{equation}
The Chandrasekhar number is a measure of the strength of the
imposed magnetic field whilst the Rayleigh number measures the
destabilising influence of the temperature gradient that is imposed
across the layer.  If all other parameters are fixed,
varying $Ra$ and $Q$ is equivalent to varying $F$ and $\kappa$. 

\begin{table}
\begin{center}
\begin{tabular}{@{}ccc}
\hline Parameter & Definition & Values used \\ \hline
$\gamma$ & $c_P/c_V$ & $5/3$\\
$m$ & $g d/R_*\Delta T - 1$ & $1.0$\\
$\theta$ & $\Delta T/T_0$ & $10.0$\\
$\kappa$ & $K/\rho_0 d c_P(R_*T_0)^{1/2}$ & $0.2$ \\
$\zeta_0$ & $\eta \rho_0 c_P /K$ & $0.1$\\
$\sigma$ & $\mu c_P/K$ & $1.0$ \\
$F$ & $B_0^2/\rho_0 \mu_0 R_*T_0$ & Variable\\
\hline
\end{tabular}
\caption{The non-dimensional parameters in the governing equations
  for compressible magnetoconvection. Note that $\rho_0$ corresponds
  to the unperturbed density at the upper surface of the layer. All
  other parameters are as defined in the text.\label{table1}}
\end{center}
\end{table}

There is a non-trivial equilibrium solution to these governing
equations, corresponding to a static polytropic layer with a uniform,
vertical magnetic field:
\begin{equation}
\mathbf{u}=0,\;T(z)=1+\theta z,\;\rho(z)=\left(1+\theta
  z\right)^m,\;\mathbf{B}=\mathbf{\hat{z}}.
\label{eq:staticstate}
\end{equation}
In this equilibrium state, the parameters that are given in
Table 1 imply that the layer of fluid is highly stratified, with the
temperature and density both varying by a factor of $11$ across the
depth of the domain.  When there is no magnetic field present (i.e. when
$Q=0$ or, equivalently, $F=0$), the critical Rayleigh number for the
onset of convection in this layer is approximately, $Ra=1189$.  For the
parameters given in Table 1, $Ra=6000$, hence this layer is
convectively unstable in the absence of an imposed magnetic field.  As
in Boussinesq magnetoconvection \citep{chand}, magnetic fields tend to
inhibit convection.  Hence, non-zero values of $Q$ lead to an increase
in the critical Rayleigh number.  Table 1 also gives values for two
diffusivity ratios, $\sigma$ and $\zeta_0$.  For simplicity, we set
$\sigma=1.0$.  The value of $\zeta_0$ plays a crucial role in
determining the near-onset bifurcation structure.  As was the case in
the truncated Boussinesq model of \citet{b99}, a choice of
$\zeta_0=0.1$ ensures that the equilibrium state can be unstable to
oscillatory modes of convection as well as stationary modes. 

\section{Numerical results}

Given the complexity of the governing equations, it is necessary to
solve these numerically.  We cover the Cartesian domain with a
computational mesh, typically consisting of $128\times 128\times 96$ grid
points.  Using standard Fast Fourier Transform (FFT) libraries, all horizontal
derivatives are evaluated in Fourier space.  Fourth-order finite
differences are used to calculate the vertical derivatives.  The
temporal evolution of this system is determined by an explicit
third-order Adams-Bashforth scheme.  This code is parallelised using MPI. 

\begin{figure}
\includegraphics[width=0.499\columnwidth]{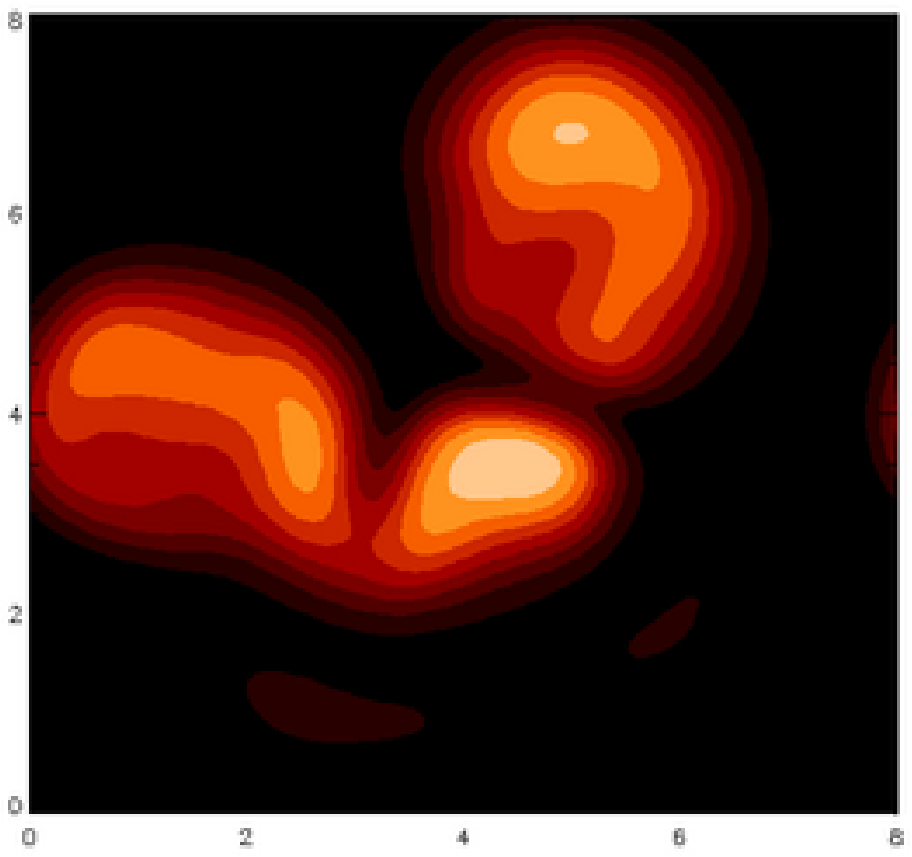}
\includegraphics[width=0.499\columnwidth]{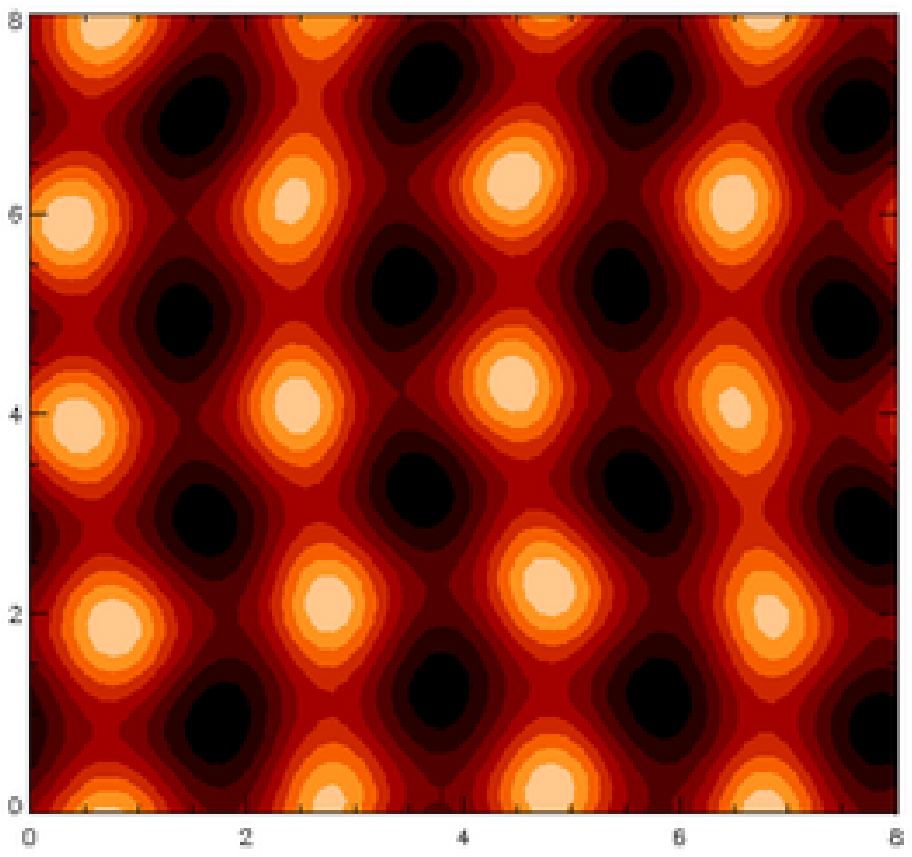}
\caption{The temperature distribution in a horizontal plane just
  below the upper surface of the computational domain for $Q=120$
  (top) and $Q=150$ (bottom). Brighter contours correspond to regions
  of warmer fluid. These simulations were both evolved from a static
  state. \label{fig1}}    
\end{figure}

\subsection{Supercritical convection}

As described in Section \ref{sec:setup}, one simple solution of these governing
equations corresponds to a static polytropic layer in the presence of
a uniform, vertical magnetic field (equation~\ref{eq:staticstate}).  We
use this static solution (plus a small thermal perturbation) as an
initial condition for the code.  For the parameter values that are
given in Table 1, this initial state is convectively unstable provided
that the Chandrasekhar number does not exceed a value of
(approximately) $Q=160$.  Since this equilibrium solution is unstable
when $Q<160$, we refer to this region of parameter space as the
`supercritical' parameter regime.

Figure~\ref{fig1} shows snapshots of the system for two different
values of $Q$ ($Q=120$ and $Q=150$), once a statistically-steady state
has been reached.  Each plot shows the temperature distribution across
a horizontal layer, just below the upper surface of the domain.  When
$Q=120$, the solution is characterised by several time-dependent
plumes from which most of the surface magnetic field has been expelled by the
convective motions.  Elsewhere the flow is dominated by magnetic
effects, so much so that the Lorentz force is strong enough to
(almost) completely inhibit convection.  Solutions of this form are
reminiscent of the `flux-separated' states that were found by
\citet{WPB} although, in that study, small-scale convective cells
tended to be found in the magnetically-dominated regions.  Here the
magnetic suppression of convection is much more pronounced in this
flux-separated state.  A very different form of solution is found when
$Q=150$ (lower part of Fig.~\ref{fig1}).  In this simulation, the
convection forms an oscillatory pattern that is distributed across the
whole of the domain.  

\begin{figure}
\includegraphics[width=\columnwidth]{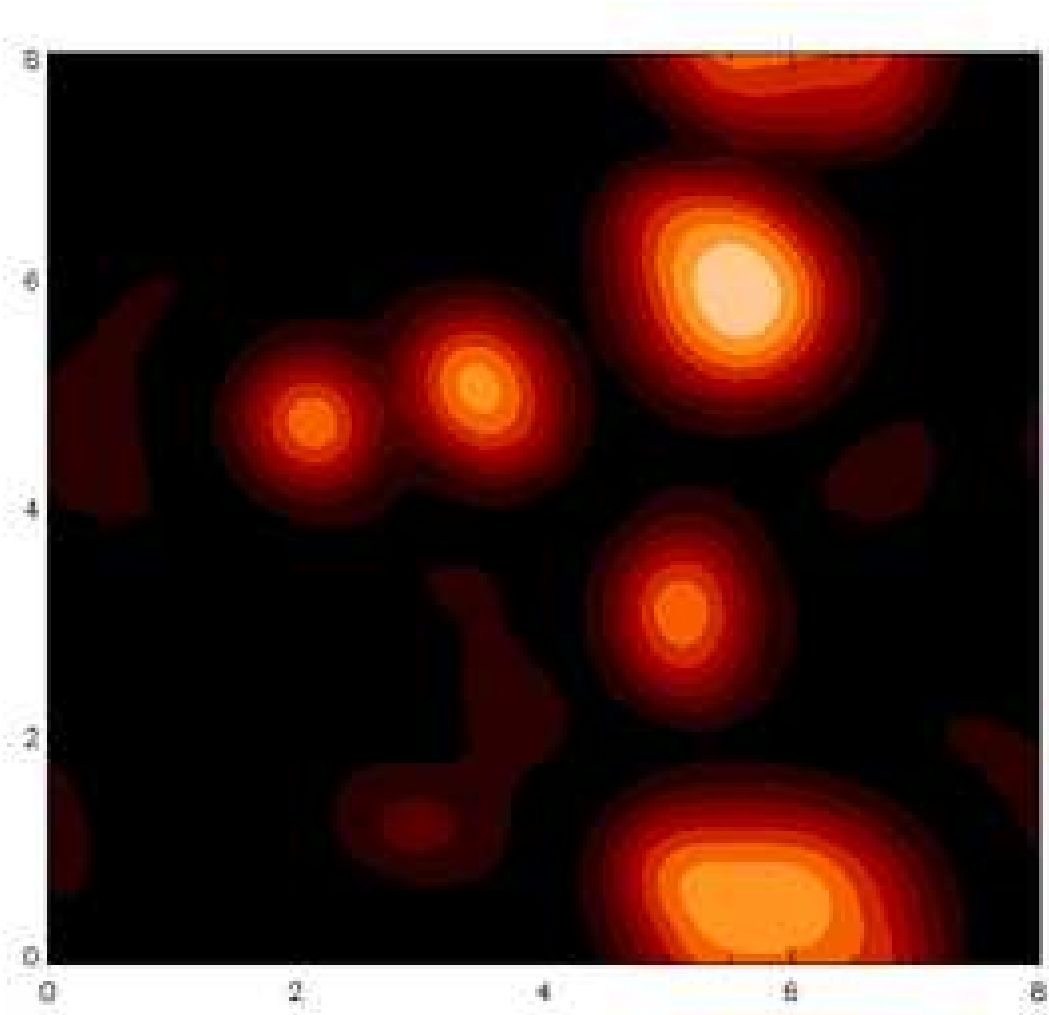}
\caption{Like the lower part of Figure~\ref{fig1}, this shows the
  temperature distribution, for $Q=150$, across a horizontal layer
  just below the upper surface of the domain. This solution has been
  generated by following the flux-separated solution
  branch.\label{fig2}}    
\end{figure}

Linear theory \citep[see, for example,][]{chand} predicts
oscillatory convection in the magnetically-dominated regime, at least
for low values of $\zeta_0$.  Hence the oscillatory behaviour of the simulation at
$Q=150$ is unsurprising.  The convection at $Q=120$ is also
time-dependent, as noted above, but this flux-separated state is representative of a 
different solution branch.  Although the present model is much more
complicated than the two-dimensional Boussinesq model that was
considered by \citet{b99}, it is clearly of interest to relate the two
studies.  The flux separated state at $Q=120$ would correspond to one
of the multiple roll states that are shown in the bifurcation diagram
in Figure 3 of \citet{b99}.  The existence of different solution
branches raises the possibility that a given set of parameter values
could be associated with more than one stable state.  Following the
procedure that is described by \citet{b99}, solution branches can be
followed by starting a simulation with fully developed convection,
rather than evolving it from a static state.  Having adjusted the value
of $Q$, the simulation can then be evolved again until a
statistically-steady state has been found.  For example, if the $Q=120$
solution is taken as an initial condition, the flux-separated solution
branch can be followed by gradually increasing the strength of the the
magnetic field (by increasing the value of $Q$).  The outcome of such
a procedure is illustrated in Figure~\ref{fig2}, which shows the
temperature distribution for a flux-separated solution at
$Q=150$.  Comparing this plot with the lower part of Figure~\ref{fig1},
it is clear that (at least) two distinct solutions exist for this set
of parameter values.  This flux-separated solution has smaller
convective plumes than the $Q=120$ case.  This is a consequence of the
fact that magnetic fields are now strong enough to inhibit convection
in a larger proportion of the computational domain.  The typical scale
of convection is reduced by increasing the strength of the imposed
magnetic field.  Additionally the rate at which convective plumes
  merge together, and split apart, reduces as the strength of the
  imposed magnetic field is increased.

\subsection{The subcritical parameter regime}

As noted above, the static polytropic layer is (linearly) stable to
convective perturbations for values of the Chandrasekhar number in
excess of approximately $Q=160$.  Hence, we refer to this range of
parameter space as the `subcritical' regime.  In order to find
non-trivial behaviour in this parameter regime, it is clearly
necessary to adopt a non-static initial condition for any simulations
that are carried out.  Given the results that were presented in the
previous section, it is natural to try to track the flux-separated
solution branch into this subcritical regime by incrementally
increasing the value of $Q$ (as before, looking for a
statistically-steady state before each increment).  By following an
analogous procedure, \citet{b99} found localised states in a
two-dimensional model, so this would appear to be the most sensible
approach.  

\begin{figure}
\includegraphics[width=\columnwidth]{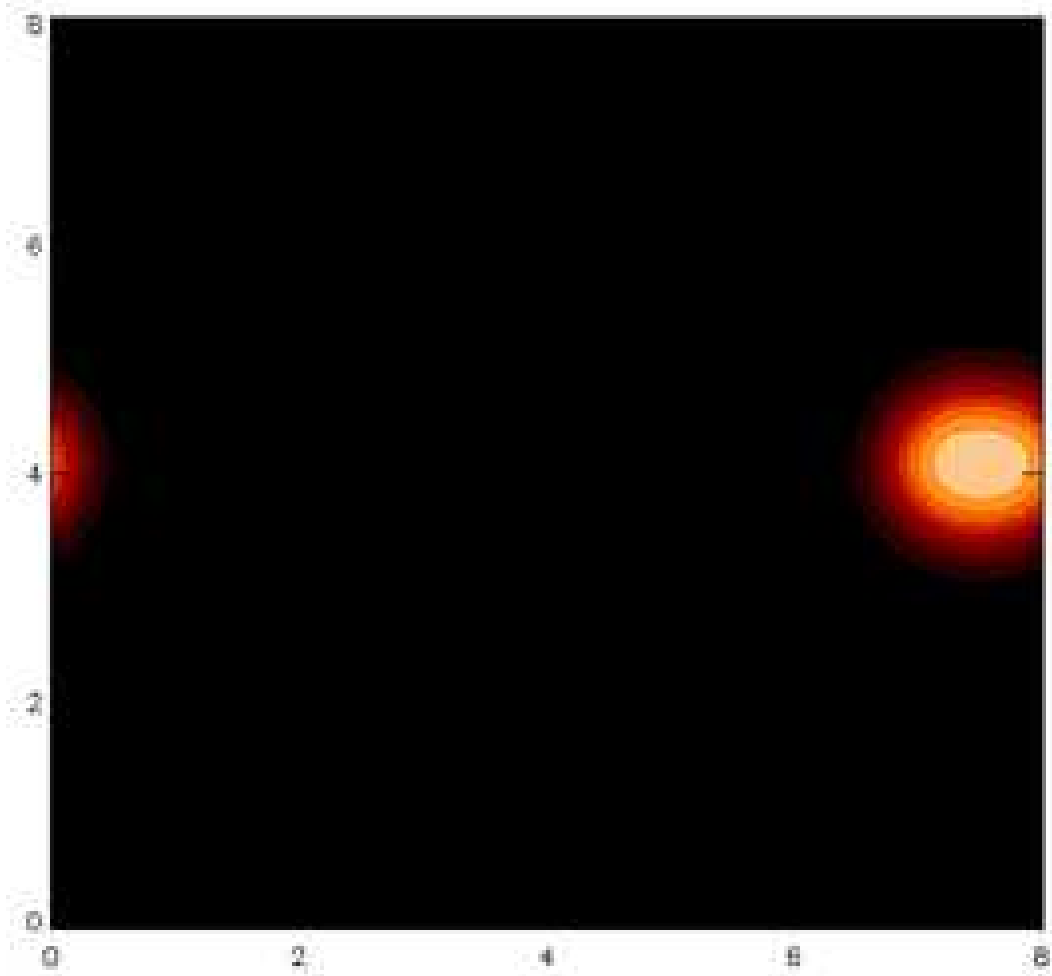}
\caption{Steady, localised convection at Q=215. This shows the
  temperature distribution in a horizontal plane just below the upper
  surface of the computational domain. Taking into account the
  periodic boundary conditions, this solution corresponds to a single
  convective plume.  The convecton is close to axisymmetric, but
  slightly elongated in the $x$-direction.}
\label{fig3}    
\end{figure}

\begin{figure}
\includegraphics[width=\columnwidth]{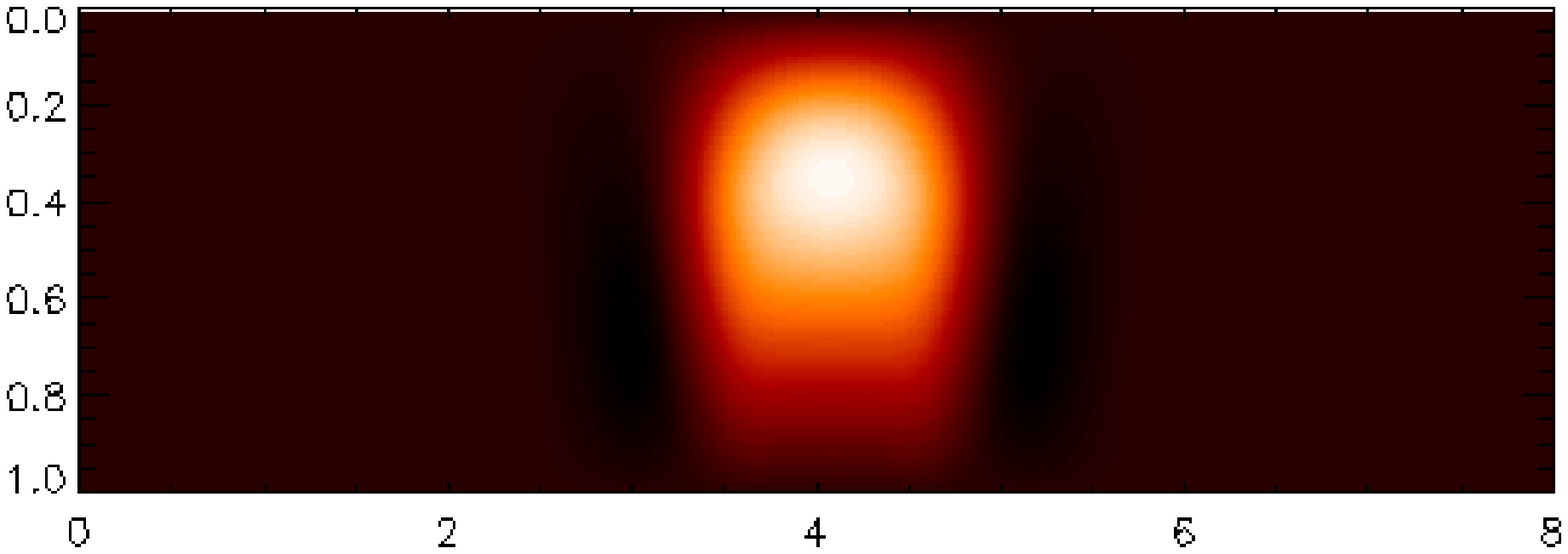}\\
\includegraphics[width=\columnwidth]{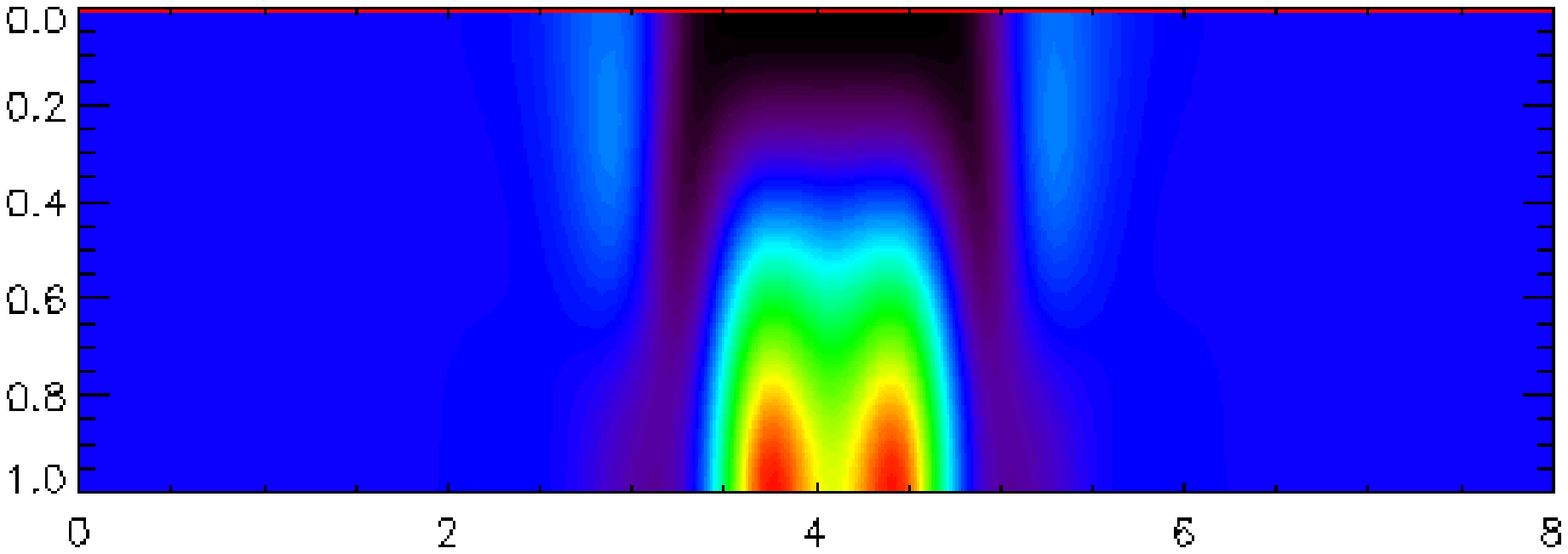}
\caption{As Figure~\ref{fig3}, these plots show snapshots of a
  convecton at $Q=215$. Top: The perturbation to the background temperature distribution along a
  vertical slice through $x=7.5$. Contours range from $-0.11$
  (black) to $0.87$ (white).
  Bottom: The vertical component  of
  the magnetic field in the same vertical slice.  Contours range
  from $0.22$ (black) to $3.67$ (red).
}    
\label{fig4}
\end{figure}

As the value of $Q$ is gradually increased, following the
flux-separated branch, the dynamical influence of the magnetic field
becomes greater.  This leads to a reduction in both the number and
scale of the field-free convective plumes.  This process continues
until a single steady plume remains.  This plume is the
three-dimensional analogue of the two-dimensional `convecton'
solutions that were found by \citet{b99}.  This three-dimensional
convecton is illustrated in Figures~\ref{fig3} and~\ref{fig4}.  At
$Q=215$, this localised convective plume is almost axisymmetric, being
slightly elongated in the $x$-direction, with a
broad central upwelling, surrounded by a narrow downflow
region.  Convection is completely suppressed everywhere else by a
strong, uniform, vertical magnetic field.  The magnetic field
distribution within the convecton is more complicated.  At the surface,
the field is almost completely expelled by the diverging convective
flows.  Towards the base of the plume, converging convective flows lead
to an accumulation of vertical magnetic flux at the base of the
convective upflow.  The stratification of the layer clearly does play a
role in determining the structure of this localised convective
feature: slightly larger temperature perturbations are found near the
top of the layer.  It should be stressed that this convecton is steady
and is therefore not simply a transient phenomenon.  It is also worth
noting that, although the horizontally averaged convective flux is
small, the local perturbations to the thermodynamic variables are of
order unity within the convecton itself.  Therefore this is also a
dynamically significant feature, albeit a highly localised one.  


It should be noted that this convecton is not truly localised
rather, by virtue of the horizontal boundary conditions, it is part of
a periodic array of such convective structures.  As seen in
figure~\ref{fig4} each structure is separated by approximately $6$
non-dimensional units which is approximately $3$ times the width of the
convective structure.  Therefore, any influence from neighbouring
convectons will be small.  An initial investigation in larger
computational domains confirms this to be the case.  Similar localised
states can be found, although the range of $Q$ for which they exist
does change with domain size.  Attempts to find an analytic
description of a perfectly axisymmetric convecton are ongoing.

\begin{figure*}
\includegraphics[width=0.4999\columnwidth]{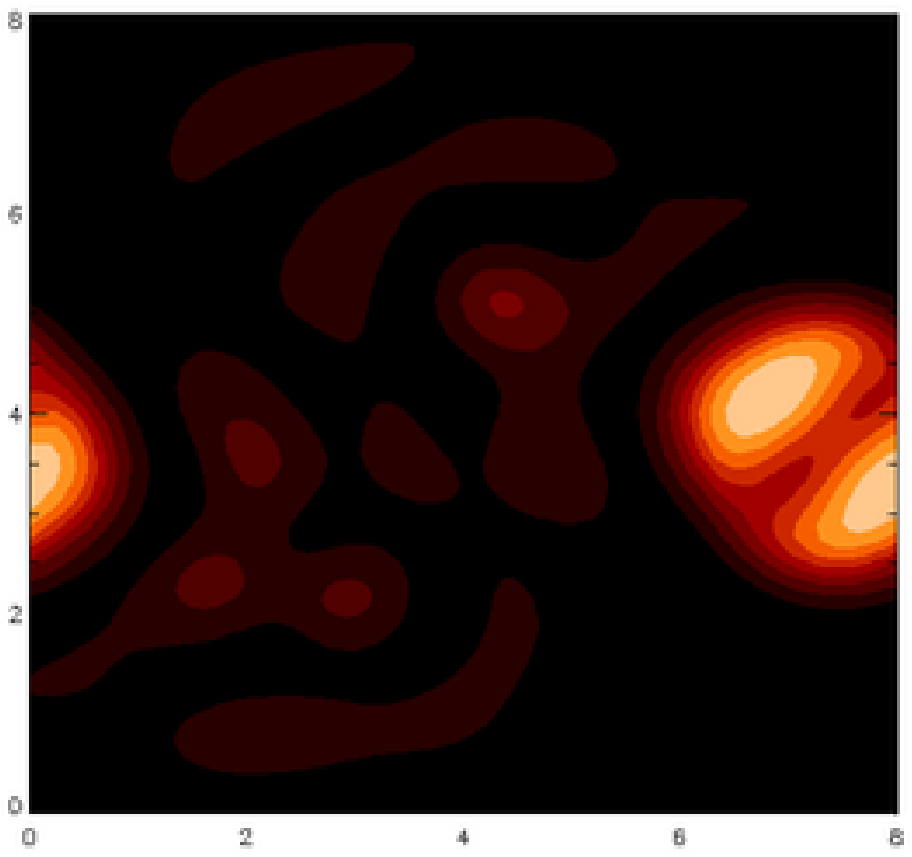}
\includegraphics[width=0.4999\columnwidth]{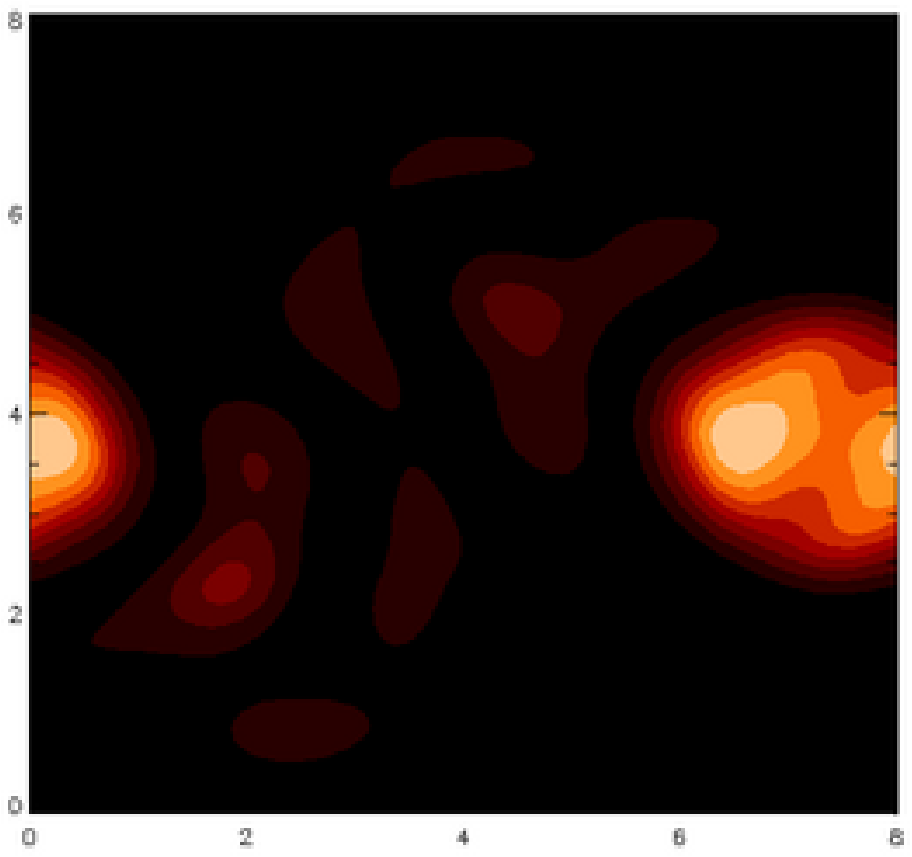}
\includegraphics[width=0.4999\columnwidth]{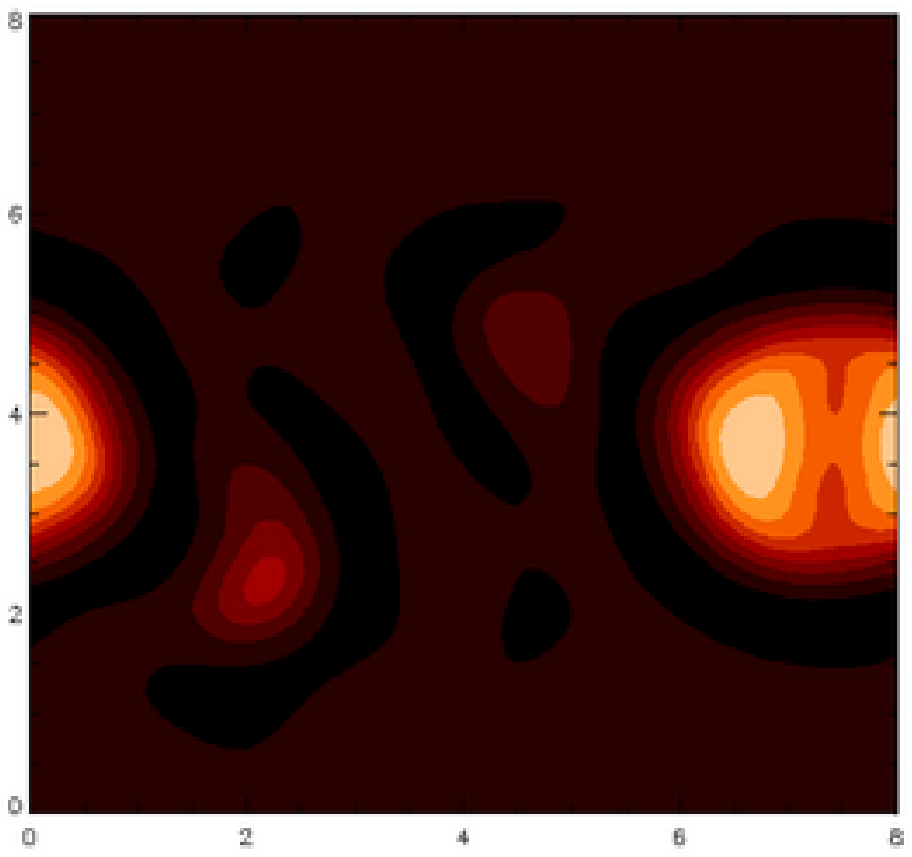}
\includegraphics[width=0.4999\columnwidth]{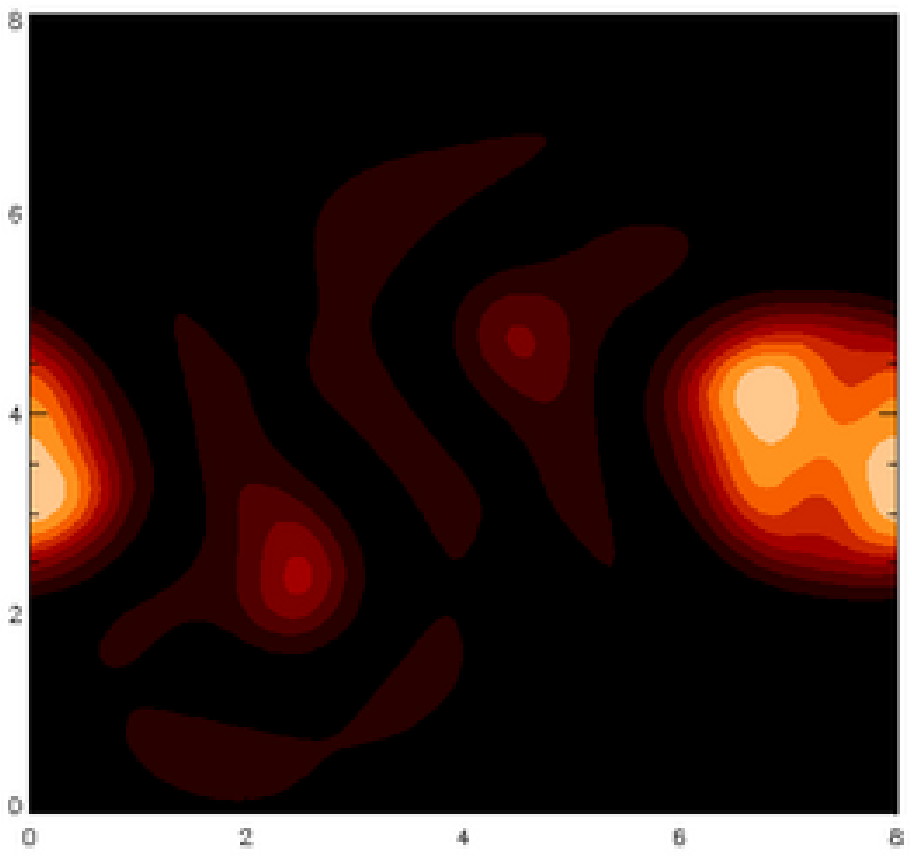}
\caption{Snapshots of convection along the convecton solution
  branch (at $Q=160$). These plots show the temperature
  distribution in a horizontal plane just below the upper surface of
  the computational domain.  These plots are separated by $4.55$ time
  units, the full cycle period is $18.20$ time units.
}
\label{fig5}
\end{figure*}

Having found this convecton, this solution branch can also be
tracked to determine its range of stability.  In fact, the solution
that is shown in Figures~\ref{fig3} and~\ref{fig4} corresponds to the
largest value of $Q$ at which the convecton is found to be
stable.  This is also the value of $Q$ at which the greatest degree of
localisation is found.  As this solution branch is followed into the
weaker field regime (by gradually \emph{decreasing} the value of $Q$),
the convecton grows, becoming increasingly asymmetric.  For example,
at $Q=190$ the solution is still highly localised and steady, but the
convective cell is about $25\%$ wider than the convecton at $Q=215$
and is more elongated in the $x$-direction.  For this larger
localised state, it is plausible that the finite size of the
computational domain is (weakly) influencing the symmetry of the
convective plume.  Reducing
$Q$ still further, we find that the solution is still mostly
localised, although there are now weak (but significant) fluctuations
elsewhere in the domain.  More importantly, the convecton is no longer
steady.  Instead, the plume `wobbles' periodically about its central
axis (as illustrated in Figure~\ref{fig5}).  This is not a true
oscillatory state, and the amplitude of the fluctuations is small, but
this behaviour is certainly interesting.  Similar vacillation has
  been seen as a way for steady convection patterns to lose stability
  via a Hopf bifurcation \citep{RWBP}.  It is also possible that the
  observed oscillations are a consequence of the finite geometry, in
  the sense that the
  convecton could be interacting with periodic copies of itself.
  Calculations in larger domains to 
investigate this further are currently underway.

\section{Discussion and Conclusions}

In these numerical simulations we have found strongly localised
convective plumes in the subcritical regime of compressible
magnetoconvection. These states are not embedded within a background
of diffuse weak convection, but rather a static fluid (see figure
\ref{fig3}). It is believed that strongly localised, steady states of this
type have not previously been observed in three spatial
dimensions. Whether umbral dots ever exhibit such an extreme degree of
localisation is unclear, but these results do suggest that highly localised
plumes could occur if the magnetic fields within sunspot umbrae
are strong enough (locally) to produce subcritical conditions for
magnetoconvection. 

The strongly localised states found in this work are found by
  continuation of flux-separated states to higher values of $Q$.  A
  similar procedure was carried out by
  \cite{WPB}, however, in their work it was found that for
  sufficiently strong imposed magnetic fields the flux separated
  solution was lost in favour of small-scale regular convection.
  \cite{WPB} used the parameter value $\zeta_0=0.2$ whereas in this
  work $\zeta_0 = 0.1$.  It is well-known that the value of $\zeta_0$
  controls the near-onset bifurcation structure.  Comparison of this
  work with the findings of \cite{WPB} indicate that it also controls
  whether or not the solution branch corresponding to flux-separated
  states continues into the subcritical regime.

Although it is appealing to relate these highly localised convective
plumes to umbral dots, one aspect of these solutions that does not
compare favourably with umbral dots is that fact that these localised
states are steady (or, at best, weakly time-dependent, as shown in
Figure~\ref{fig5}). So, unlike the observed umbral dots, these
solutions do not have finite lifetimes. This may be associated with
the simplified nature of the model, which assumes a uniform background
state: spatiotemporal inhomogeneities in the umbral background may
play an important role in determining the lifetime of an umbral
dot. Alternatively, time-dependent behaviour may be the result of
interacting plumes. Perhaps the time-dependent state that is shown in
Figure~\ref{fig2} is a more realistic representation of the
distribution of convective plumes within sunspot umbrae. Work is
ongoing to establish whether or not it is possible to find truly
localised states that exhibit significant time-dependence. The
simplified model of \citet{bw02} suggests that oscillatory solutions
should exist, however, while we cannot rule out this possibility,
oscillatory localised states have not yet been observed in the full
system of equations.   

This model is clearly a highly idealised representation of
magnetoconvection within sunspot umbrae. Truly localised states have
not yet been found in more realistic models of photospheric
magnetoconvection. It would be of great interest to establish whether
or not processes such as radiative transfer promote (or inhibit) the
formation of strongly localised plumes. It should also be stressed
that the range of parameter values that can be considered in numerical
models of this type bear little resemblance to true photospheric
values (although this is true of all numerical magnetoconvection
calculations, regardless of the level of physical complexity). Having
said that, we believe that results from models of this type do provide
some useful insights into photospheric magnetoconvection. 

Magnetoconvection is not the only branch of physics in which strongly localised
states are relevant. Similar states have now been found in a
variety of other systems, including binary fluid convection
\citep{bk08}, buckling rod problems, nonlinear optics \citep{vlad02}
and experiments on a ferrofluid in an applied magnetic field
\citep{cake05}. In a more abstract setting, localised states have also
been found in the one-dimensional bistable Swift-Hohenberg
equation \citep{bk06,bk07a,bk07b}.  The results presented in this
paper share many similarities with the behaviour of the
Swift-Hohenberg equation in one dimension, as well as the other physical
systems mentioned above.  There is now a good theoretical
understanding of localised states in one extended dimension, however,
in higher dimensions the problem is not well understood, and has
recently been posed as an open problem \citep{k08}. Having said that,
some recent progress has been made. For example, \citet{lloyd} have
investigated the properties of spatially localised states in the
two-dimensional Swift-Hohenberg equation. In future work, we intend to
further explore the relationship between the behaviour of this simple
pattern forming system and the magnetoconvection equations.  

{\bf Acknowledgements:}
This work has been supported by EPSRC grant EP/D032334/1 (SMH) and
STFC grant ST/H002332/1.  PJB would like to acknowledge the support
of STFC.  Large-scale computations were performed on the UKMHD
Consortium machine based in St. Andrews, UK.

\bibliographystyle{mn2e}
\bibliography{biblio}

\begin{thebibliography}{}

\bibitem[\protect\citeauthoryear{Bergeon \& Knobloch}{Bergeon \&
  Knobloch}{2008}]{bk08}
Bergeon A.,  Knobloch E.,  2008, Phys. Fluids., 20, 034102

\bibitem[\protect\citeauthoryear{Bharti, Beeck \& {Sch\"ussler}}{Bharti
  et~al.}{2010}]{barti10}
Bharti L.,  Beeck B.,    {Sch\"ussler} M.,  2010, A\&A, 510, A12

\bibitem[\protect\citeauthoryear{Bharti, Jain \& Jaaffrey}{Bharti
  et~al.}{2007}]{bjj07}
Bharti L.,  Jain R.,    Jaaffrey S.~N.~A.,  2007, ApJ, 665, L79

\bibitem[\protect\citeauthoryear{Blanchflower}{Blanchflower}{1999}]{b99}
Blanchflower S.,  1999, Phys. Lett. A, 261, 74

\bibitem[\protect\citeauthoryear{Blanchflower \& Weiss}{Blanchflower \&
  Weiss}{2002}]{bw02}
Blanchflower S.~M.,  Weiss N.~O.,  2002, Phys. Lett. A, 294, 297

\bibitem[\protect\citeauthoryear{Burke \& Knobloch}{Burke \&
  Knobloch}{2006}]{bk06}
Burke J.,  Knobloch E.,  2006, Phys. Rev. E, 73, 056211

\bibitem[\protect\citeauthoryear{Burke \& Knobloch}{Burke \&
  Knobloch}{2007a}]{bk07a}
Burke J.,  Knobloch E.,  2007a, Chaos, 17, 037102

\bibitem[\protect\citeauthoryear{Burke \& Knobloch}{Burke \&
  Knobloch}{2007b}]{bk07b}
Burke J.,  Knobloch E.,  2007b, Phys. Lett. A, 360, 681

\bibitem[\protect\citeauthoryear{Bushby \& Houghton}{Bushby \&
  Houghton}{2005}]{bh05}
Bushby P.~J.,  Houghton S.~M.,  2005, MNRAS, 362, 313

\bibitem[\protect\citeauthoryear{Chandrasekhar}{Chandrasekhar}{1961}]{chand}
Chandrasekhar S.,  1961, Hydrodynamic and Hydromagnetic Stability.
Dover, New York

\bibitem[\protect\citeauthoryear{Danielson}{Danielson}{1964}]{dan64}
Danielson R.~E.,  1964, ApJ, 139, 45

\bibitem[\protect\citeauthoryear{Dawes}{Dawes}{2007}]{d07}
Dawes J.~H.~P.,  2007, Journal of Fluid Mechanics, 570, 385

\bibitem[\protect\citeauthoryear{Deinzer}{Deinzer}{1965}]{dein65}
Deinzer W.,  1965, ApJ, 141, 548

\bibitem[\protect\citeauthoryear{{Kitai} et~al.,}{{Kitai}
  et~al.}{2007}]{Kitai}
{Kitai} R.,  et~al., 2007, PASJ, 59, S585

\bibitem[\protect\citeauthoryear{Knobloch}{Knobloch}{2008}]{k08}
Knobloch E.,  2008, Nonlinearity, 21, T45

\bibitem[\protect\citeauthoryear{Lloyd, Sandstede, Avitabile \&
  Champneys}{Lloyd et~al.}{2008}]{lloyd}
Lloyd D.~J.~B.,  Sandstede B.,  Avitabile D.,    Champneys A.~R.,  2008, SIAM
  Journal on Applied Dynamical Systems, 7, 1049

\bibitem[\protect\citeauthoryear{Matthews, Proctor \& Weiss}{Matthews
  et~al.}{1995}]{matt}
Matthews P.~C.,  Proctor M.~R.~E.,    Weiss N.~O.,  1995, Journal of Fluid
  Mechanics, 305, 281

\bibitem[\protect\citeauthoryear{Ortiz, Rubio \& van~der Voort}{Ortiz
  et~al.}{2010}]{ort10}
Ortiz A.,  Rubio L.~R.~B.,    van~der Voort L.~R.,  2010, ApJ, 713, 1282

\bibitem[\protect\citeauthoryear{Richter \& Barashenkov}{Richter \&
  Barashenkov}{2005}]{cake05}
Richter R.,  Barashenkov I.~V.,  2005, Phys. Rev. Lett., 94, 184503

\bibitem[\protect\citeauthoryear{Rucklidge, Weiss, Brownjohn \&
  Proctor}{Rucklidge et~al.}{1993}]{RWBP}
Rucklidge A.~M.,  Weiss N.~O.,  Brownjohn D.~P.,    Proctor M.~R.~E.,  1993,
  Geophys. Astrophys. Fluid Dynamics, 68, 133

\bibitem[\protect\citeauthoryear{{Sch\"ussler} \& {V\"ogler}}{{Sch\"ussler} \&
  {V\"ogler}}{2006}]{sv06}
{Sch\"ussler} M.,  {V\"ogler} A.,  2006, ApJ, 641, L73

\bibitem[\protect\citeauthoryear{{Sobotka}, {Bonet}, {Vazquez} \&
  {Hanslmeier}}{{Sobotka} et~al.}{1995}]{sob95}
{Sobotka} M.,  {Bonet} J.~A.,  {Vazquez} M.,    {Hanslmeier} A.,  1995, ApJ,
  447, L133

\bibitem[\protect\citeauthoryear{Sobotka, Brandt \& Simon}{Sobotka
  et~al.}{1997}]{sob97}
Sobotka M.,  Brandt P.~N.,    Simon G.~W.,  1997, A\&A, 328, 682

\bibitem[\protect\citeauthoryear{Sobotka \& Hanslmeier}{Sobotka \&
  Hanslmeier}{2005}]{sob05}
Sobotka M.,  Hanslmeier A.,  2005, A\&A, 442, 323

\bibitem[\protect\citeauthoryear{{Socas-Navarro}, {Mart{\'{\i}}nez Pillet},
  {Sobotka} \& {V{\'a}zquez}}{{Socas-Navarro} et~al.}{2004}]{Socas}
{Socas-Navarro} H.,  {Mart{\'{\i}}nez Pillet} V.,  {Sobotka} M.,
  {V{\'a}zquez} M.,  2004, ApJ, 614, 448

\bibitem[\protect\citeauthoryear{Tao, Weiss, Brownjohn \& Proctor}{Tao
  et~al.}{1998}]{twbp98}
Tao L.,  Weiss N.~O.,  Brownjohn D.~P.,    Proctor M.~R.~E.,  1998, ApJ, p.~L39

\bibitem[\protect\citeauthoryear{{Thomas} \& {Weiss}}{{Thomas} \&
  {Weiss}}{2008}]{TW}
{Thomas} J.~H.,  {Weiss} N.~O.,  2008, {Sunspots and Starspots}.
Cambridge University Press

\bibitem[\protect\citeauthoryear{Vladimirov, {McSloy}, Skryabin \&
  Firth}{Vladimirov et~al.}{2002}]{vlad02}
Vladimirov A.~G.,  {McSloy} J.~M.,  Skryabin D.~V.,    Firth W.~J.,  2002,
  Phys. Rev. E, 65, 046606

\bibitem[\protect\citeauthoryear{Weiss}{Weiss}{1966}]{w66}
Weiss N.~O.,  1966, Proc. Roy. Soc. Lond. A, 293, 310

\bibitem[\protect\citeauthoryear{Weiss, Proctor \& Brownjohn}{Weiss
  et~al.}{2002}]{WPB}
Weiss N.~O.,  Proctor M.~R.~E.,    Brownjohn D.~P.,  2002, MNRAS, 337, 293

\end{thebibliography}

\label{lastpage}

\end{document}